\documentclass[]{aastex631}
\usepackage{graphicx} % Required for inserting images
\usepackage{amsmath}

\begin{document}
\title{The recent anomalously weak polar field does not imply a weak field at solar cycle 25 minimum}
%\maketitle

\correspondingauthor{Zi-Fan Wang}
\email{zfwang@nao.cas.cn}

\author{Zi-Fan Wang}
\affiliation{State Key Laboratory of Solar Activity and Space Weather, National Astronomical Observatories, Chinese Academy of Sciences, Beijing 100101, China}
\affiliation{School of Astronomy and Space Science, University of Chinese Academy of Sciences, Beijing, China}

\author[0000-0001-5002-0577]{Jie Jiang}
\affiliation{School of Space and Earth Sciences, Beihang University, Beijing, People’s Republic of China}
\affiliation{Key Laboratory of Space Environment Monitoring and Information Processing of MIIT, Beijing, People’s Republic of China}

\author{Yukun Luo}
\affiliation{School of Physical Science and Technology, Southwest Jiaotong University, Chengdu 611756, People’s Republic of China}
\affiliation{Astrophysical Center, Southwest Jiaotong University, Chengdu 611756, People’s Republic of China}

\author{Ruihui Wang}
\affiliation{School of Space and Earth Sciences, Beihang University, Beijing, People’s Republic of China}
\affiliation{Key Laboratory of Space Environment Monitoring and Information Processing of MIIT, Beijing, People’s Republic of China}

\author{Xinchun Ma}
\affiliation{School of Space and Earth Sciences, Beihang University, Beijing, People’s Republic of China}
\affiliation{Key Laboratory of Space Environment Monitoring and Information Processing of MIIT, Beijing, People’s Republic of China}

\author{Jing-Xiu Wang}

\affiliation{State Key Laboratory of Solar Activity and Space Weather, National Astronomical Observatories, Chinese Academy of Sciences, Beijing 100101, China}
\affiliation{School of Astronomy and Space Science, University of Chinese Academy of Sciences, Beijing, China}

\begin{abstract}
    The ongoing solar cycle 25 has progressed past its peak of sunspot numbers, being stronger than the previous cycle 24.  However, the present polar field is rather weak compared to previous cycles at the same evolution phase, particularly in the northern hemisphere, where it has been decreasing since mid-2025 till present.  A prominent poleward surge is observed to cause the decrease of the polar field.  This raises concerns to the polar field at cycle 25 minimum, which is the precursor to the strength of the next cycle 26.  To predict whether the polar field at cycle 25 minimum will be weak as expected, we use observation-based statistical properties to predict the active region emergence during the latter half of cycle 25, and use a surface flux transport model to simulate the evolution of the large-scale magnetic field.  We predict the polar field at cycle 25 minimum to be $-5.62\pm1.61$~G in the north and $5.51\pm1.48$~G in the south, both stronger than those at cycle 24 minimum. This is because the poleward surge causing the temporal decrease of the polar field originates from a group of active regions that produces net increase to the polar field, instead of active regions with non-Joy's tilt.  Our results suggest the weak polar field at present is unlikely to cause a weak minium and an exceptionally weak cycle 26, clarifying that the short term evolution should not be simply correlated to the long term properties of the solar large-scale field.

%    The ongoing solar cycle 25 has been active since its rising phase, and has reached its peak of sunspot numbers, revealed to be stronger than the previous cycle 24.  However, recent studies suggest that its polar field during year 2026 is weak compared to solar cycles of the same evolution phase as a result of a poleward surge of the leading polarity, raising concern on the polar field at the cycle minimum, which is crucial to the evolution of the next cycle 26 suggested by dynamo models.  We use observation-based statistical properties to predict the AR emergence during the latter half of cycle 25, and use a surface flux transport model to simulate the evolution of the large-scale magnetic field.  We predict the polar field at cycle 25 minimum to be $-5.78\pm1.67$~G and $5.69\pm1.77$~G for both hemispheres, stronger than that of cycle 24 minimum. The active regions contributing to the poleward surge that temporary reducing the polar field in 2026 actually have net positive contributions to the polar field at cycle minimum instead of rogue active regions with extreme properties.  Our results show that the short term evolution of the polar field should not be simply correlated to the cycle minimum, and that cycle 26 will unlikely to be exceptionally weak.
\end{abstract}

\section{Introduction}
%%CJS2016, hathaway2016

%emphasis on the surge, and descriptions of ARs

The current solar cycle 25 has just past its peak of sunspot numbers \citep{2025RNAAS...9..202G}.  During its rising phase, cycle 25 has been active, with solar activity triggering geomagnetic storms \citep{2024JASS...41..171K,2025ApJ...982..194P}, which are of vital importance to further understanding of space weather and to the security of the technological society in the future.  Statistically, solar cycle 25 is stronger than the weakest cycle within century, i.e., cycle 24, from the perspective of various solar and geomagnetic properties \citep{2025EGUGA..27.2858H,2025ApJ...990L..55J,2026Ge&Ae.tmp....1A}.  This suggests that the Sun has reversed the trend of decreasing cycle strength since cycle 21, instead of entering a Maunder-like minimum \citep{2025ApJ...990L..55J}.
%Solar cycle is one of the most important features of solar activity, its prediction is of value to physical understanding of the Sun and to\

\begin{figure}
\gridline{\fig{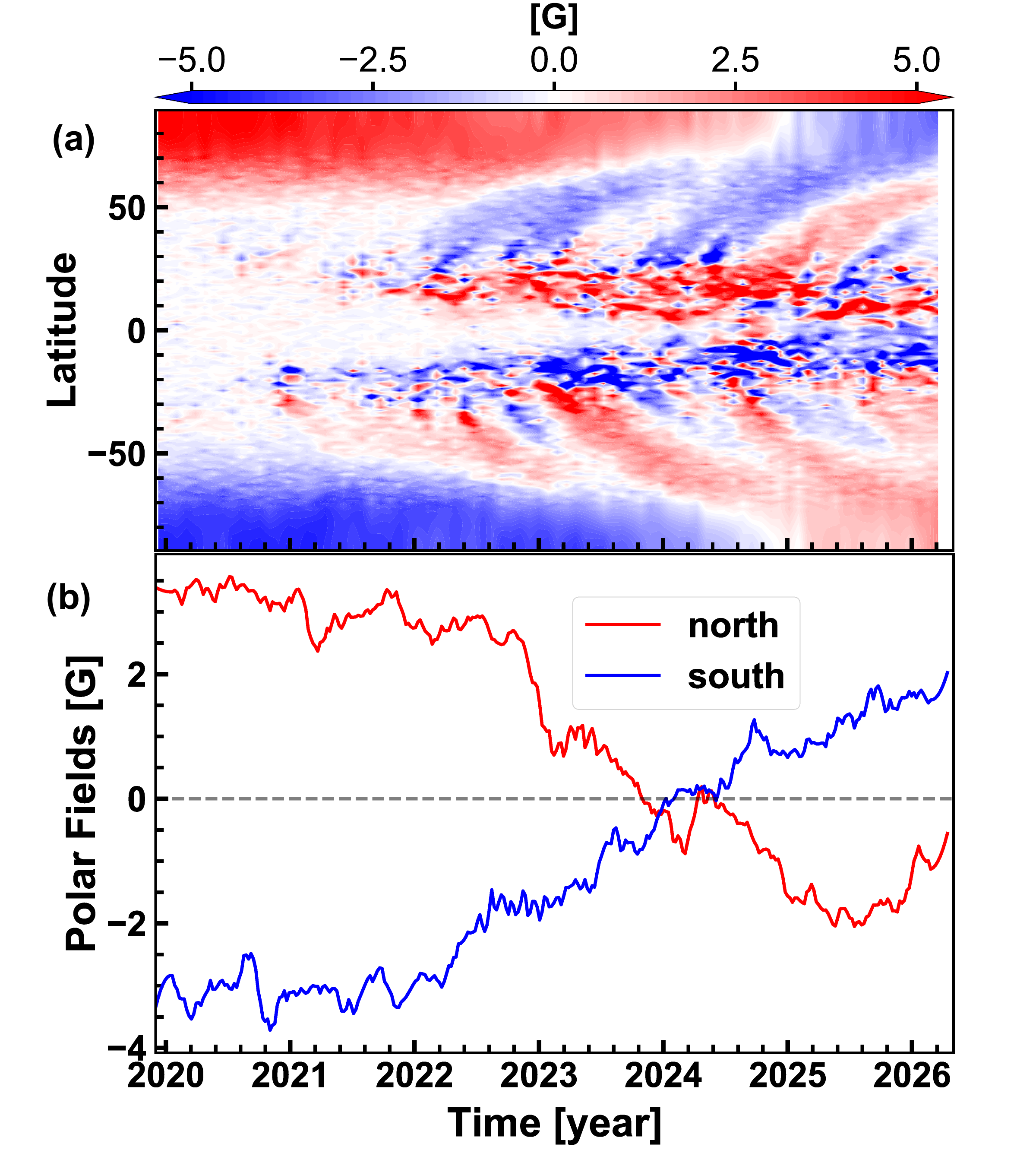}{0.6\textwidth}{}          }
\caption{Observational large-scale magnetic field evolution of the Sun during cycle 25.  Panel (a) shows the latitudinally averaged time-latitude diagram (i.e., magnetic butterfly diagram), from the data of HMI.  Panel (b) shows the evolution of the polar field from the data of HMI, obtained by averaging the magnetic field between latitudes $60^\circ$ and $90^\circ$, and is smoothed by averaging within a window of 3 months.  Red curve indicates the north pole while blue indicates the south pole. \label{fig:polarobs}}
\end{figure}

However, the polar field observed at present is quite weak, despite the solar activity.  As shown by the latitudinally averaged magnetic field (i.e., magnetic butterfly diagram) by the Helioseismic and Magnetic Imager of the Solar Dynamics Observatory (SDO/HMI, \citet{2012SoPh..275..207S}) in Figure \ref{fig:polarobs}(a), the polar field reverses during year 2024.  After approximately 1 year of the polar field reversal, the polar field at the northern hemisphere begins to decrease till present, shown in Figure \ref{fig:polarobs}(b).  \citet{2026ApJ..1000..197W} has shown that the current polar field, especially at the northern hemisphere, is weaker than several previous cycles, and has predicted that the polar field to remain weak up to October of year 2026.  The polar field is crucial to the evolution of solar cycles as the polar field at cycle minimum is positively correlated to the strength of the next cycle \citep{1978GeoRL...5..411S,2005GeoRL..32.1104S,2005GeoRL..3221106S,2007MNRAS.381.1527J,2007PhRvL..98m1103C}, known as the polar field precursor.  In solar dynamo models involving the mutual generation of poloidal and toroidal fields \citep{1955ApJ...122..293P},  this is interpreted as that the generation of toroidal field from poloidal field is mostly linear.  If the current status of the polar field leads to a weak polar field at cycle minimum, cycle 26 will then be notably weak. This urges the realistic prediction of cycle 25 minimum, and on the understanding of the current weakening of the polar field.

%However, does cycle 25 being stronger than 24 mean that the solar cycle strength will continue to grow during the upcoming cycles?  Recently, \citet{2026ApJ..1000..197W} has revealed that the polar field of cycle 25 up to October of year 2026 will be fairly weak compared to cycles 21-24.  The observed polar field evolution during the early months of 2026 up to April shows a clear weakening trend in the northern hemisphere, as shown in Figure \ref{fig:polarobs}.  If this trend persists to the end of cycle 25, it will lead to a relatively weak cycle 26.  This is because the polar field at cycle minimum is positively correlated to the strength of the next cycle \citep{1978GeoRL...5..411S,2005GeoRL..32.1104S,2005GeoRL..3221106S,2007MNRAS.381.1527J,2007PhRvL..98m1103C}, known as the polar field precursor.  In solar dynamo models involving the mutual generation of poloidal and toroidal fields \citep{1955ApJ...122..293P},  this is interpreted as that the generation of toroidal field from poloidal field is mostly linear.  The current status of cycle 25 raises concerns on its minimum and the next cycle 26.

The evolution of the polar field is contributed by the active regions (ARs) emerged during the cycle, according to the \citet{Babcock1961}-\citet{Leighton1969} mechanism.  ARs are generally formed by the rise and emergence of toroidal field, composed of a pair of opposite polarities with a tilt angle to the latitudinal direction.  As a result of the surface flux transport (SFT) process, part of the AR magnetic flux migrates poleward, contributing to the polar field evolution.  Then, the polar field  at the cycle minimum determines the evolution of the next solar cycle.  Observational and numerical studies of the SFT model such as \citet{1985AuJPh..38..999D,1989ApJ...347..529W,1998ApJ...501..866V,2002SoPh..209..287M,2014ApJ...791....5J} has been concretizing the B-L mechanism, leading to the the polar field, both short term evolution and the cycle minimum, well explained by the properties of ARs.

Just as it can be seen in Figure \ref{fig:polarobs}(a), the poleward transportation of AR flux is not uniform in time, but in the form of poleward surges \citep{1981SoPh...74..131H}, stretching from the activity latitudes toward the poles, which can affect the short term evolution of the polar field.  Especially, a poleward surge of the leading polarity (colored red in the northern hemisphere) originates from the activity latitudes during late 2024, and migrates poleward during 2025, and is responsible for the decrease of the northern polar field as shown by \citet{2026ApJ..1000..197W}.  Typically, an AR with tilt angle following Joy's law \citep{Hale1919} can form a poleward surge of the following polarity, and another more diffusive surge of the leading polarity \citep{2002SoPh..209..287M,2015SoPh..290.3189Y,2019ApJ...871...16J,2025ApJ...987....1W}, and both surges will be more intense if the emerging latitude is higher.  On the other hand, the contribution of AR to the polar field at the cycle minimum is larger for lower emerging latitudes and larger tilt angle \citep{2014ApJ...791....5J,Petrovay2020JSWSC}.  Moreover, the most prominent surges may be a result of a series of continuously emerging ARs \citep{2020ApJ...904...62W}.  Hence, the origin of the poleward surge during 2025 and its influence to the polar field should not be simply determined and requires more analysis.

A realistic prediction to the polar field at cycle minimum requires prediction of the AR emergence throughout the remaining cycle, combined with SFT model.
%Meanwhile, if the AR emergence till the end of the ongoing cycle can be realistically predicted, the polar field at cycle 25 minimum can be predicted, leading to the prediction of cycle 26. 
During the declining phase of cycle 24, \citet{2016ApJ...823L..22C} utilized the observed statistical AR properties by \citet{1994SoPh..151..177H,2010ApJ...719..264C,2011A&A...528A..82J} for SFT simulations, obtained the polar field at cycle minimum, and predicted a moderate cycle 25.  The uncertainty of the prediction comes from the uncertainty of AR properties.  This SFT-precursor method is further developed and evaluated by \citet{2018ApJ...863..159J}.  Especially, the authors show that the error of predicting the cycle profile (usually expressed by sunspot number) is stabilized after about 3 years has passed for the current cycle.  The uncertainty of the next cycle decreases as more of the current cycle is known.  Considering this, an SFT-precursor prediction with uncertainty for cycle 26 is practical now. 

In this letter, we predict the emergence of ARs of the latter half of solar cycle 25, and use an SFT model to predict the evolution of the large-scale field evolution till the end of cycle 25.  We will show that the strength of the polar field at cycle 25 minimum is likely to be stronger than cycle 24 minimum despite the temporal weakening of the polar field during year 2025, which will probably lead to a moderate cycle 26.  We also provide an analysis of the origin and influence of the temporal weakening of the polar field during year 2026, by evaluating the AR emergence and SFT simulation.% We will then show that the poleward flux surge of the leading polarity causing the temporal weakening is produced by normal ARs instead of rogue ones, hence having limited influence to the polar field at the end of the cycle.

%In this letter, we present the SFT-precursor prediction of cycle 26 based on realistic observational AR properties.  We will demonstrate that cycle 26 is equally possible to be either weaker or stronger than cycle 25.  %We will then show that rogue active regions occurring during (\textbf{time}) are responsible for the weak cycle 26.%provide an explanation that specific active regions during cycle 25 contributes to the weak cycle 26.

\section{Methods}
\subsection{Prediction of active region emergence}\label{subsec:ARs}

The SFT-precursor method is composed of the prediction of AR emergence during the ongoing cycle based on already know observations and statistical properties of ARs from previous cycles, and inserting the predicted regions into the SFT model along with an initial magnetic field to obtain the polar field at cycle minimum.  A number of random realizations of AR predictions are made and simulated in order to obtain the uncertainty range of solar cycle prediction.

The method of predicting AR emergence in this letter is adopted from \citet{2018ApJ...863..159J} and \citet{Jiang2020}, with a few ingredients changed according to newer observations.  We summarize the method below.

\begin{figure}

\gridline{\fig{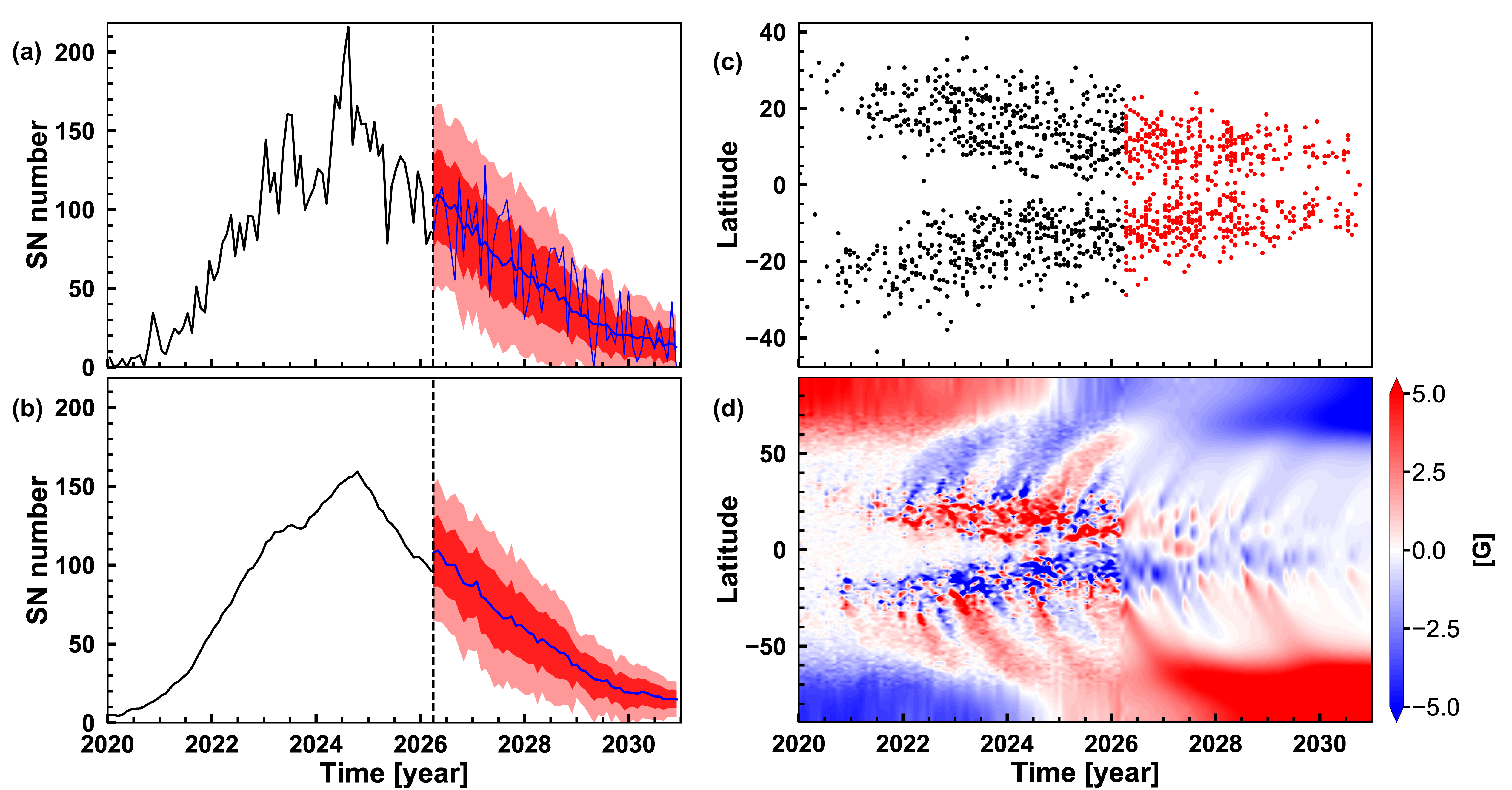}{1\textwidth}{}          }
\caption{Statistical properties of AR emergence prediction and one example of random realizations.  Panel (a) shows the monthly AR number, with black curve representing observations, thick blue curve representing the average of all random realizations, thin blue curve representing one random realization, deep and light red shades representing 1 and 2 standard deviation ranges.  Panel (b) is similar to (a), showing the 13-month smoothed version of the AR number, and the blue curve representing average of all random realizations.  Panel (c) shows the butterfly diagram of one random realizations, with black representing observational ARs and red representing predicted ARs.  Panel (d) shows the magnetic butterfly diagram as a result of the evolution of the ARs shown in panel (c).\label{fig:pre_ars}}
\end{figure}

The emergence rate of ARs is depicted by the monthly sunspot numbers (SN).  The profile of SN is proposed by the equation 1 of \citet{1994SoPh..151..177H}, which describes the ascending phase of solar cycles with good accuracy, due to the well-known Waldmeier effect \citep{1955epds.book.....W}.  The declining phase often deviate from the function, which is demonstrated and corrected by \citet{2018ApJ...863..159J}. By fitting to cycles 12$\sim$24, the time dependent standard deviation of SN is obtained.  Then, the known half of cycle 25 is fitted to the function of SN, and time dependent random component is added according to the standard deviation.  The mean of all random realizations and the standard deviation ranges are shown in Figures \ref{fig:pre_ars}(a) and (b).

The bipolar ARs are added according to the monthly SN obtained above, with exact daily emergence being random while satisfying the total monthly number.  The ARs are attributed with a Gaussian distributed latitude centered around the mean latitude dependent on the phase of cycle \citep{2011A&A...528A..82J}.  Such dependence generates a typical butterfly diagram, shown in Figure \ref{fig:pre_ars}(c).  The mean latitude is positively correlated to cycle amplitude as shown by \citet{Jiang2020}.  The ARs are then attributed with random hemisphere and longitude.  The area of ARs are determined by the distribution of Equation 12 and 13 in \citet{2011A&A...528A..82J}.

For the tilt angle, the Joy's law is known to have large scatter \citep{2014ApJ...791....5J} as well as systematic negative correlation to cycle strength \citep{2010A&A...518A...7D}, with the latter known as tilt quenching.  In this latter we adopt newer observational results for tilt scatter and tilt quenching.  We choose a linear form of the Joy's law, $\alpha=T_n|\lambda|+\sigma\left(A\right)\cdot X$, in which $\alpha$ denotes the tilt angle, $T_n$ denotes the tilt coefficent dependent on the strength of cycle n, $\sigma\left(A\right)$ is the tilt scatter dependent on area A, and X represents a random variable following normal distribution.  The tilt quenching is obtained from \citet{2021A&A...653A..27J} as $T_n=-0.00107S_n+0.61$, where $S_n$ is the amplitude of cycle n.

The tilt scatter is obtained from \citet{2025ApJ...986..114Q}'s mutual validation tilt angle data set covering newest observations of cycles 23 and 24.  In this letter, we make a further refinement to the results, by subtracting the contribution of Joy's law from the tilt angle values, before calculating the tilt scatter.  The tilt scatter is $\sigma\left(A\right)=-2.72A+21.09$, in which area is in the unit of millionth of hemisphere.  After adding the scatter, the tilt angle is further multiplied with a factor of 0.7, as a result of the effect of inflow toward activity belts, as it is introduced and calibrated in \citet{2010ApJ...719..264C}, and later used in \citet{2014ApJ...791....5J,2018ApJ...863..159J,Jiang2020}.  The magnetic flux of ARs is also necessary for the bipolar AR input of the SFT model, which is adopted from \citet{2025ApJ...986..114Q} as $F=10^{20.88}A^{0.57}$, in which the total flux F is in Maxwells.

%The properties of active regions are mostly adopted from Jiang 2020 and 2018.

%The distribution of AR flux and tilt scatter is from Qin2025.

%When calculating the tilt scatter, the Joy's law itself is subtracted from the original tilt angle values before calculating the scatter, resulting a different scatter-area relation.

%Figure 1: tilt scatter as a function of area

%Figure 1: prediction of active region emergence: panel a, unsmoothed SN; panel b, smoothed SN, panel c. example of a butterfly diagram

%\begin{figure}

%\gridline{\fig{fig2.pdf}{0.6\textwidth}{}          }
%\caption{Prediction of AR emergence.  Panel (a) shows the monthly AR number, with black curve representing observations, thick blue curve representing the average of all random realizations, thin blue curve representing one random realization, deep and light red shades representing 1 and 2 standard deviation ranges.  Panel (b) is similar to (b), showing the 13-month smoothed version of the AR number, and the blue curve representing average of all random realizations.  Panel (c) shows the butterfly diagram of one random realizations, with black representing observational ARs and red representing predicted ARs.\label{fig:pre_ars}}
%\end{figure}

\subsection{The surface flux transport model}

%Should be the same version used in recent works...?
The SFT model is governed by the radial component of the induction equation limited on the photosphere surface.  The equation in spherical coordinates ($r,\theta,\phi$), is as follows,  
\begin{equation}\label{SFTeq}
\begin{aligned}
  \frac{\partial B_{r}}{\partial t}=-\Omega \left ( \theta  \right )\frac{\partial B_{r}}{\partial \phi }-\frac{1}{R_{\odot } \sin \theta }\frac{\partial }{\partial \theta }\left [v_{\theta}\left ( \theta  \right ) B_{r}\sin \left ( \theta  \right ) \right ]+\\
  \frac{\eta_{T} }{R{_{\odot }}^{2}}\left [\frac{1}{\sin \theta } \frac{\partial }{\partial \theta }\left ( \sin \theta \frac{\partial B_{r}}{\partial \theta } \right ) +\frac{1}{\sin ^{2}\theta }\frac{\partial ^{2}B_{r}}{\partial \phi^{2}}\right ]+\\
  S\left ( \theta ,\phi ,t \right ),
\end{aligned}
\end{equation}
in which $B_r$ is the radial component of the magnetic field, $R_{\odot }$ is the radius of the Sun, $\Omega \left ( \theta  \right )$ is the differential rotation at the surface, $v_{\theta}\left ( \theta  \right )$ is the meridional flow at the surface, $\eta_T$ is the turbulent diffusion coefficient, and $S\left ( \theta ,\phi ,t \right )$ is the source term contributed by the emergence of ARs.

%The advective and diffusive terms follow those of \citet{2026ApJ..1000..197W}.  

The differential rotation is adopted from \citet{1983ApJ...270..288S}.  For the meridional flow, the function from \citet{2026ApJ..1000..197W} is used, with a peak velocity of $13m~s^{-1}$.  The turbulent diffusion coefficient is set as $450km^{2}~s^{-1}$.

The source terms is composed of bipolar ARs generated according to the location, time, tilt angle, and flux obtained in Subsection \ref{subsec:ARs}.  Each AR has two Gaussian magnetic flux patches with opposite polarity.  The latitude and longitude of the two polarities are determined by the center of the AR, the tilt angle, and the angular separation of the bipole in the form $\Delta\beta=0.45\sqrt{A_R}$, in which $A_R$ is the sum of sunspot group area and facular area given by \citet{1997ApJ...482..541C}.  The profile of the flux patches is as follows,

\begin{equation}
    B_{r,\pm}\left(\theta,\phi\right)=\pm B_{max}\exp\left(\frac{2\left(1-\cos{\beta_\pm\left(\theta,\phi\right)}\right)}{\delta^2}\right),
\end{equation}
in which $\beta_\pm\left(\theta,\phi\right)$ is the angular distance of point $\left(\theta,\phi\right)$ to the center of the flux patch, and $\delta=0.4\Delta\beta$ is the width of the flux patch.  

%Generally, the following polarity is know to be more diffusive than the leading polarity, which affects the SFT evolution results of ARs \citep{2019ApJ...883...24I,2020SoPh..295..119Y,2021A&A...650A..87W}.  Hence, we introduce an asymmetry factor $f_{\pm}$, so the width of the flux patch is determined as $\delta_\pm=0.4f_\pm \Delta \beta$.  The asymmetry factor is $f_+=1,f_-=\sqrt{2}$.  Then, the maximum field strength $B_{max}$ is determined so that the total flux matches the flux values given in Subsection \ref{subsec:ARs}.

The bipolar ARs are inserted into the SFT model on their corresponding day.  The initial magnetic field is the synoptic magnetogram of vertical magnetic field at the photosphere during Carrington rotation (CR) 2309 (April of year 2026), from HMI.  The SFT equation is solved by a spectral method code identical to \citet{2025ApJ...993...27L}.

\section{Results}
\subsection{Prediction of the large-scale field evolution of cycle 25}

We run 100 simulations to obtain the expectation of cycle prediction and the uncertainty.  The simulations start at year 2026 and ends at 2031.  We express the results in term of the evolution of the polar field and the axial dipole moment.  The polar field is defined as the field strength averaged between latitudes $60^\circ$ and $75^\circ$.  The axial dipole moment is an equivalent expression of the polar field, which reads,

\begin{equation}
    D=\frac{3}{4\pi}\int_0^{2\pi}d\phi\int_0^\pi B\left(\theta,\phi\right)\cos{\theta}\sin{\theta}d\theta
\end{equation}

\begin{figure}

\gridline{\fig{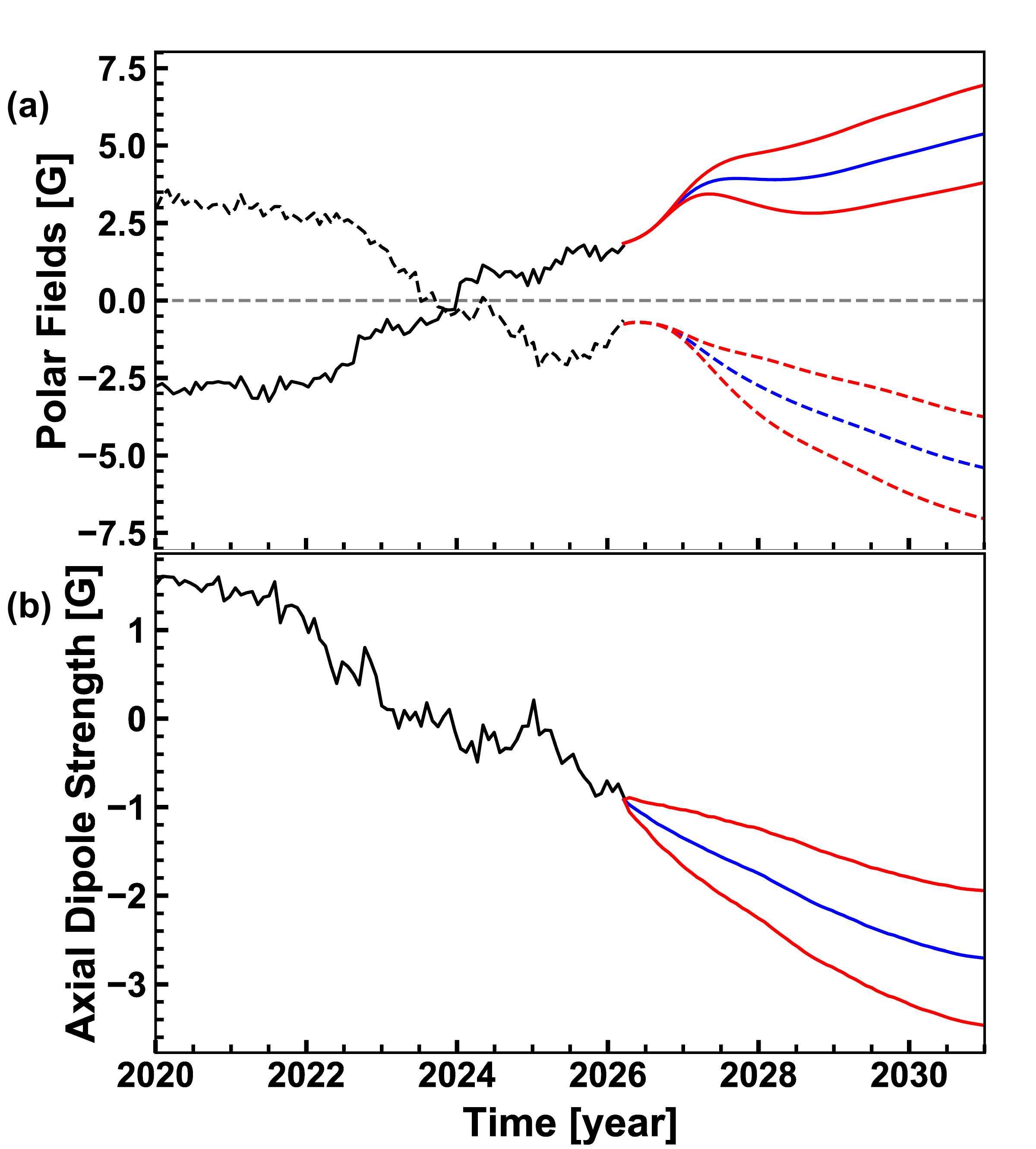}{0.6\textwidth}{}          }
\caption{Prediction of solar cycle evolution.  Panel (a) shows the evolution of the polar field, with black curve representing observations from HMI, blue solid and dashed curves representing the average values of predictions for the southern and northern hemispheres, respectively, and the red curves representing the corresponding 1 standard deviation ranges.  Panel (b) shows the evolution of the axial dipole moment, with black representing observations from HMI, blue representing average of predictions, and red representing 1 standard deviation ranges.\label{fig:pre_cyc}}
\end{figure}

%The results are presented in Figure 2.  As shown, 
During the first year of the simulation, the polar field evolution is mainly determined by the preexisting fields in the initial magnetic field.  The significant poleward surge originating in year 2025 and reaching the pole in 2026, which was suggested to be the cause of temporary polar field decrease in the north pole by \citet{2026ApJ..1000..197W}, is also shown in the magnetic butterfly diagram in Figure \ref{fig:pre_ars}(d).  The predicted polar field evolution during the first year of the simulation, shown in Figure \ref{fig:pre_cyc}(a), is consistent to \citet{2026ApJ..1000..197W}, and has negligible uncertainty from new AR emergence compared to later part of the simulation.   Since the end of year 2026, the mean values of the polar field and axial dipole moment gradually increases, and their uncertainty increases as well due to the randomness of ARs, until the end of cycle 25.   

%The magnetic butterfly diagram in Figure \ref{fig:pre_cyc}(a) shows a significant poleward surge of the leading polarity in the northern hemisphere originated in about 2025, reaching the pole during 2026.  This corresponds to a temporary decrease in polar field strength shown in Figure \ref{fig:pre_cyc}(b), during which the uncertainty from new AR emergence is negligible.  Since the end of year 2026, the uncertainty of polar field and axial dipole moment gradually increases as a result of the randomness of ARs, until the end of cycle 25. 

At the end of cycle 25, the polar fields at the northern and southern hemisphere are $-5.62\pm1.61$~G and $5.51\pm1.48$~G, respectively, and the axial dipole moment is $-2.70\pm0.76$~G, as shown in Figures \ref{fig:pre_cyc}(a) and (b).  The values are obtained from the average within 7 CRs.  The uncertainty ranges here are 1 standard deviation of all the random realizations.  The predicted polar fields are of $1\sigma$ stronger than those at the end of cycle 24, which are 2.5 G; the same is true for the axial dipole moment.   \citet{2026ApJ..1000..197W} have shown that there exists a large difference between solar cycles on how much the first 6 year of a cycle can contribute to the final polar field at the cycle minimum.  Our simulation here shows clearly that while the polar fields in 2026 are similar in strength to cycle 24 of the same phase, the final field strength is actually predicted to be stronger, which is in agreement with the former concept.

%Using the polar field precursor relationship used by \citet{2018ApJ...863..159J}, i.e., $S_{n+1}=58.7D_n$, the amplitude of cycle 26 is $166.5\pm47.5$.  According to the 13-month smoothed International Sunspot Number data, the ongoing cycle 25 peaks at about 160, so the difference between cycle 25 and the predicted expectation of cycle 26 (about $\sim6$) is much smaller than the uncertainty range $47.5$.  From this perspective, there is a even chance for cycle 26 to be either stronger or weaker than 25.

%\subsection{Rogue active regions causing the weak polar field}
\subsection{Source of the poleward surge reducing polar field during year 2026}
%(the name ``rogue active regions" has been used by several papers and I think we can simply use this fancy name as well -- just to attract people)

\begin{figure}

\gridline{\fig{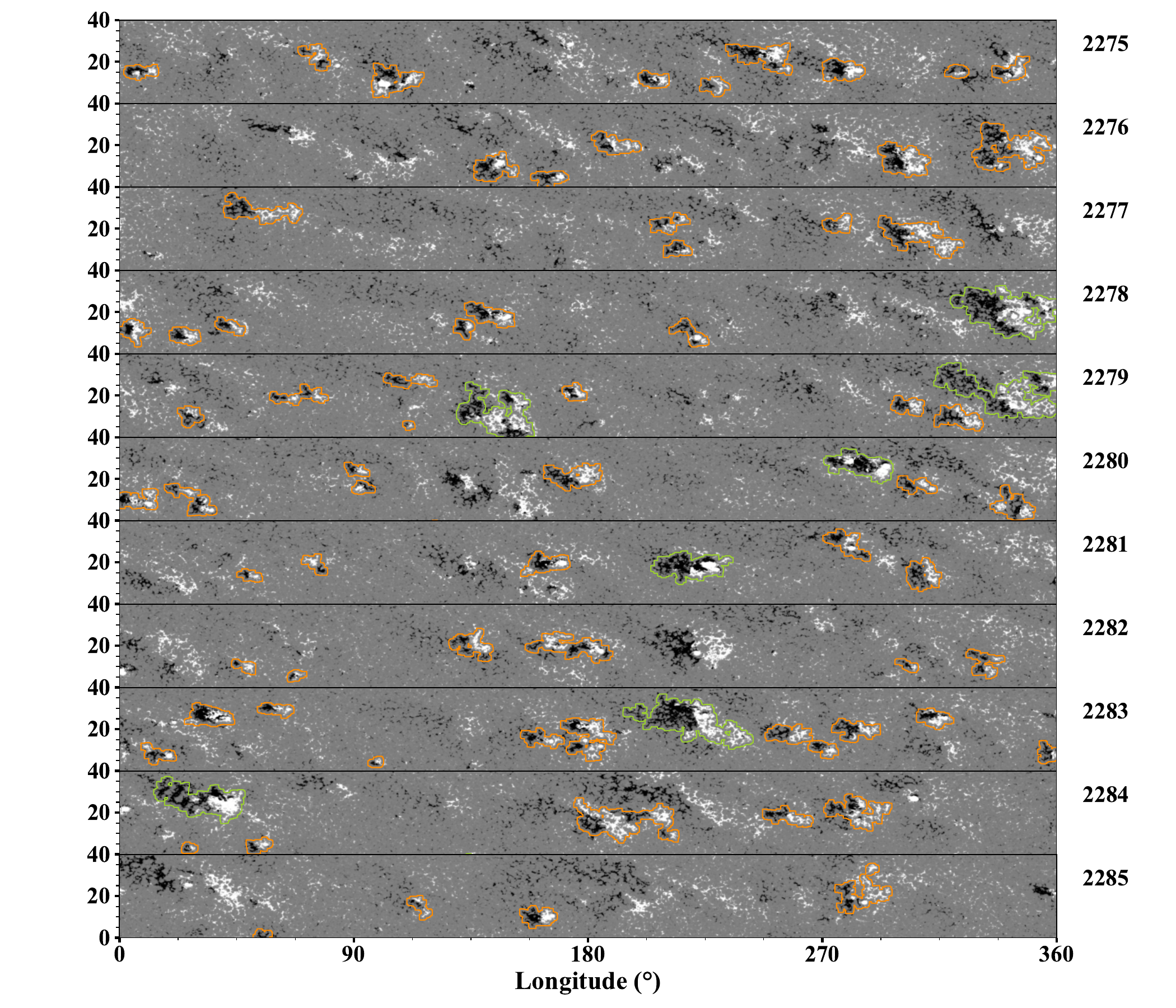}{0.9\textwidth}{}          }
\caption{Stack plot displaying ARs during CRs 2275-2285.  Each tile is obtained from the synoptic maps for photospheric radial magnetic field of HMI, containing only latitudes from $0^{\circ}$ to $40^\circ$ in the northern hemisphere, stacked from the top to bottom by chronological order.  White and black indicate positive and negative fields, respectively, and the colors saturate at 100G.  ARs from the database of \citet{2023ApJS..268...55W,2024ApJ...971..110W} are shown by contours.  Orange contours mark ARs with unsigned flux less than $4\times 10^{22}$~Mx, while green contours mark ARs with unsigned flux greater than $4\times 10^{22}$~Mx.\label{fig:stack}}
\end{figure}

Since the polar field at the cycle minimum of cycle 25 is not likely to be as weak as it appears in year 2026, then comes the question that what is the nature of the temporal weakening of the polar field during 2026 as well as the related poleward surge.  To analyze this, we evaluate the properties of the originating ARs.  Judging from the magnetic butterfly diagram Figure \ref{fig:polarobs}(a), we focus on the ARs emerging on the northern hemisphere during CRs 2275-2285 (September 2023 to June 2024).  The ARs are obtained from the AR database by \citet{2023ApJS..268...55W,2024ApJ...971..110W} by auto detection from HMI.  We display these ARs by stacking synoptic maps and marking them in the stack plot, in Figure \ref{fig:stack}.  Especially, ARs with total flux larger than $4\times 10^{22}$~Mx are marked with green contours.

Being close to the maximum of cycle 25, CRs 2275-2285 are characterized by extensive emergence of ARs.  Some ARs tend to emerge in proximity to each other, during the same CR, or during several consecutive CRs, forming activity complexes or nests, for example, the ARs emerging during CRs 2275-2279, within longitudes $315^\circ$ to $360^\circ$.  The large AR marked green during CR 2278 actually corresponds to several NOAA AR numbers, i.e., 13490, 13491, 13492, 13495, 13497, and 13502.  There are also large ARs that emerge in a relatively isolated manner and last for several CRs, for example, the large AR marked green during CR 2281, which corresponds to NOAA AR 13590.

Some ARs during CRs 2275-2285, especially those larger ones marked green, lie around latitude of $20^\circ$, which is considered high latitude from the perspective of contributions of ARs to the final polar field and dipole moment at cycle minimum, according to the results of \citet{2014ApJ...791....5J,Petrovay2020}.  Being closer to the pole, such ARs can form prominent poleward surges of the following polarity and then the leading polarity, given that they have Joy's tilt.  When the surges reach the pole, they cause rapid temporal evolution of the pole consisting of increase and then decrease.  On the other hand, their contribution to the final polar field is limited, due to the cancellation of the two polarities \citep{2014ApJ...791....5J,2015SoPh..290.3189Y,Petrovay2020}.   %The dipole moment from these ARs tend to decrease gradually since the ARs enter the decay phase, and a notable portion of both the leading and following polarities migrates poleward.  Such ARs can cause rapid increase followed by decrease of polar field when the surges reach the poles.

%The database contains the contribution to the final dipole moment at the cycle minimum for each AR, denoted as $D_f$, using the method by \citet{2021A&A...650A..87W}.

%The $D_f$ of the ARs is shown in the histogram in Figure \ref{fig:df}(a).  As shown, the distribution of $D_f$ follows a distribution peaked at slightly negative values, with width much narrower compared to the distribution of all ARs during solar cycles shown in \citet{2024ApJ...971..110W}, and also has much fewer occurrence of outliers.  This suggests that the ARs we consider should have constructive contribution to the polar field at cycle minimum in general.  If we sum up the $D_f$ values within each bin of Figure \ref{fig:df}(a), we get Figure \ref{fig:df}(b), which shows that the total contribution to the final dipole is roughly evenly distributed within each bin without preferring a certain $D_f$ range.  The total final dipole moment is contributed by a group of varied ARs.

%for both hemispheres follows a distribution peaked at slightly negative values, with width much narrower compared to the distribution of all ARs during solar cycles shown in \citet{2024ApJ...971..110W}, and also has much fewer occurrence of outliers.  The distribution of the northern hemisphere is even narrower and has even fewer outliers compared to the southern hemisphere.  Hence, the ARs during CRs 2275-2285 have weak constructive contribution to the polar field at cycle minimum in general.

\begin{figure}
\gridline{\fig{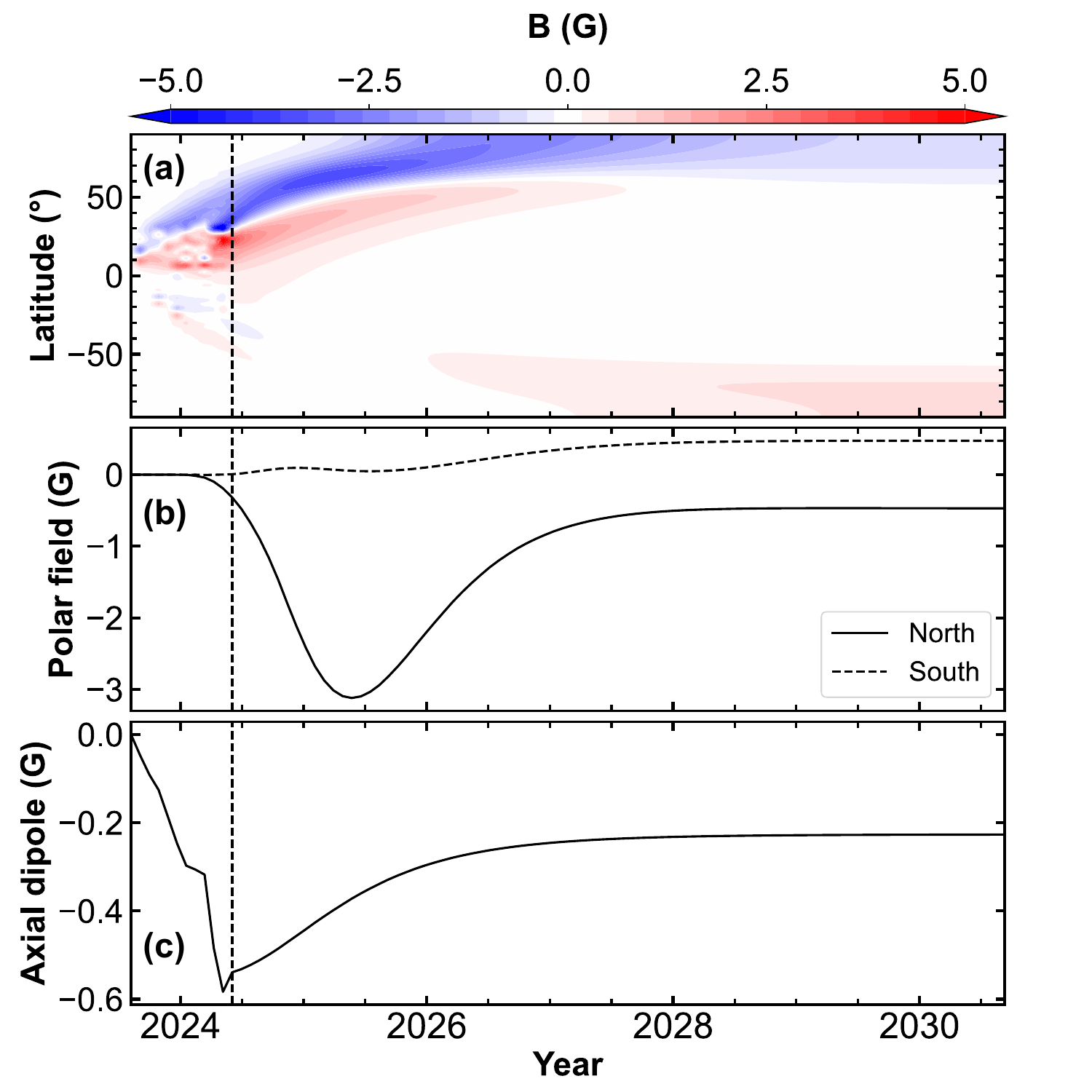}{0.6\textwidth}{}          }
\caption{SFT simulation of the ARs on the northen hemisphere during CRs 2275-2285.  Panel (a) shows the magnetic butterfly diagram.  Panel (b) shows the evolution of the polar field, with solid and dashed curves representing southern and northern hemispheres, respectively.  Panel (c) shows the evolution of the axial dipole moment.  The vertical dashed lines indicate the time when the last AR is inserted into the simulation.\label{fig:surgesim}}
\end{figure}

In order to explicitly show the evolution of these ARs, we perform an SFT simulation with all ARs on the northern hemisphere during CRs 2275-2285, without initial magnetic field, while inserting the ARs into the simulation with their realistic magnetic field configurations on their corresponding day.  The simulation produces a clear poleward surge of the leading polarity after a poleward surge of the following polarity in the northern hemisphere, as shown in Figure \ref{fig:surgesim}.  The polar field at the northern hemisphere rises at first, and then decreases during year 2026 when the surge of the leading polarity reaches the pole, but the polar field contribution is not totally negated.  The axial dipole moment increases when new ARs are being put into the simulation, and drops afterwards, yet the total dipole moment provides constructive contribution to the dipole at the cycle minimum.  The temporary rapid evolution of the polar field is characteristic to large ARs emerging at relatively high latitudes as discussed above, while the contribution of ARs to the final dipole moment is the collective contribution of all the ARs during the time considered, especially the lower-latitude ARs.  Short-term temporal evolution of the polar field and the final polar field, while both important perspectives of the large-scale field evolution of the Sun, are affected by different ARs.  The weakening of the polar field during year 2026 is part of the short-term evolution governed by high-latitude ARs, not necessarily related to the final polar field at the cycle minimum. 

%The polar field at both hemispheres rises at first, and then decreases during year 2026, but the total contribution to the polar field at cycle minimum is still constructive.  The axial dipole moment increase during the time when new ARs are being put into the simulation, and drops afterwards, yet the total dipole moment provides constructive contribution to the dipole at the cycle minimum.  From these results, we can conclude that the poleward surge of the leading polarity only causes the polar field to decrease temporarily, but the net contribution from its originating ARs is still constructive to the polar field at cycle minimum.  Therefore, weak polar field during year 2026 does is not likely to make the polar field at cycle minimum weak compared to previous cycles.

%(How many influential rogue ARs are present? Do we need a table to demonstrate their properties?)

%Figure 3: Examples of rogue active regions

%Table 1: perhaps needed?

%\begin{figure}

%\gridline{\fig{dfn.pdf}{1\textwidth}{}          }
%\caption{Contribution of ARs on the northen hemipshere during CRs 2275-2285 to the final dipole moment at cycle minimum, $D_f$.  Panel (a) shows the histogram, while panel (b) shows the sum of $D_f$ within each bin of ARs in panel (a). Note that negative values of $D_f$ suggest constructive contribution to cycle 25 minimum, considering the polarity. \label{fig:df}}
%\end{figure}

\section{Discussions and conclusions}
In this letter, we have made a prediction of the AR emergence during the latter half of solar cycle 25 based on observational empirical properties, and used them to simulate the evolution of the large-scale magnetic field with an SFT model.  We predict the polar fields at cycle minimum to be $-5.62\pm1.61$~G and $5.51\pm1.48$~G, which are of $1\sigma$ stronger than the end of cycle 24, despite the temporary weakening of polar fields during year 2026.  The poleward surge of the leading polarity contributing to the weakening of the polar field during year 2026 is part of the temporary effect by a group of high-latitude ARs during CRs 2275-2285, while the total contribution of all the ARs during the time period still increases the final polar field at cycle minimum. Our results answer to the rising concern of the minimum of cycle 25 and the following cycle 26 which is unlikely to be exceptionally weak. %The poleward surge of the leading polarity contributing to the weakening of the polar field is mostly generated by a group of ARs which increase the final polar fields at cycle minimum collectively, instead of weakening them.  Our results answer to the rising concern of the minimum of cycle 25 and the following cycle 26 which is unlikely to be exceptionally weak.  The temporal weakening of the polar field during year 2026 is a result of a group of ARs that still produces constructive contributions to the cycle minimum instead of some ARs with extreme properties.

%and predicted the amplitude of cycle 26 to be $166.5\pm47.5$.  The mean value of the predicted cycle 26 is very close to cycle 25, compared to the uncertainty range, so the probability of cycle 26 to be stronger or weaker than 25 is mostly the same.  Hence, there is no clear ``rising trend'' in the long-term solar cycle evolution beyond 1 cycle, from the current knowledge.
%Our results show that cycle 26 is probably weaker than 25, and also has a fair chance to be weaker than cycle 24, becoming the weakest among the century.

The polar field at the cycle minimum is key to cycle prediction.  If we adopt the linear relationship by \citet{2018ApJ...863..159J}, i.e., $S_{n+1}=58.7D_n$, the amplitude of cycle 26 is $158\pm45$.  According to the 13-month smoothed International Sunspot Number data, the ongoing cycle 25 peaks 160.9, so the difference between cycle 25 and the predicted expectation of cycle 26 is much smaller than the uncertainty range.  From this perspective, there is a even chance for cycle 26 to be either stronger or weaker than 25.  
However, this prediction to cycle 26 does not put the uncertainty of the precursor method itself into account yet.  Precursor method utilizing polar field and dipole moment is known to have uncertainties, and observational limits are an important cause.  
Observational limitations include the inaccuracy of high latitude observations, and the limited number of observed cycles.
%Polar field observations are inaccurate due to high latitudes, and different sources of observations may differ.  
%For example, \citet{2023JGRA..12831681U} shows that the dipole at cycle 24 minimum observed by Wilcox Solar Observatory is $20\%$ larger than that by HMI, and also an offset in time between the two dipole profiles.  
%The limited number of solar cycles with polar field observations available also limits the reliability of the precursor method.  
Especially, the current cycle 25 is stronger than several precursor method predictions using observations during the minimum of cycle 24, such as \citet{2018SoPh..293..112P,2021JSWSC..11....3J,2021ApJ...909...87K,2023JGRA..12831681U}, even with their corresponding uncertainty ranges considered (see \citet{2026RvMPP..10...11K} for a comprehensive review).   The polar field precursor, along with the process of toroidal field production it implies, should continue to be evaluated and improved in the future.  In this letter, we focus our results on the minimum of cycle 25, while the prediction of cycle 26 is a preliminary estimation. %In this letter, we provide an approximated estimation of cycle 26 that it will be similar to cycle 25.

%Caution should be taken for this prediction, as the polar field at cycle 25 minimum is notably stronger than cycle 24 minimum.  From this perspective, the polar field precursor relationship may also need future refinement.

The SFT-precursor method relies on the prediction method of AR emergence, which is adopted from \citet{Jiang2018JASTP}.  We only add one modification to the model, i.e., the tilt scatter from \citet{2025ApJ...986..114Q}. There are other recent results on the statistics of ARs that could affect the solar cycle evolution.  For example, \citet{2021ApJ...920...31M,2026A&A...710A.188N} show that the tilt angle follows t distribution with heavier tails, which may increase the variation of dipole moment produced.  Meanwhile, \citet{2026A&A...710A.188N} also shows that the pole separation scales logarithmically to AR flux.  Incorporating those results into our model requires more evaluations and numerical experiments, which will be carried out in future studies.

%However, their results do not show difference of Joy's law between cycles, which is an important ingredient in our model, i.e., latitude and tilt quenching \citep{Jiang2020}.  From this perspective, direct incorporating the results into our model may be difficult.  Further statistical evaluations of observations and improvement to the model are needed in the future.}

%\textit{In fact, the bipole separation in the model may not necessarily be what it is in the observations -- the model itself is an empirical model, and the parameters are control parameters to ensure that $D_fBMR$ is close to, at least statistically, $D_fAR$.  We all know that real ARs can be very different to BMRs, so even if we use the most accurate observational statistics for the ideal, symmetric BMRs, we may not get a realistic result -- we still need to optimize the parameters so that the dipole production of BMRs are at least reasonably match those ARs in reality.}

The SFT-precursor prediction of solar cycles is based on the assumption that the solar cycle is primarily determined by its previous one, instead of earlier ones, which is in agreement with a number of previous statistical and physical models \citep{1978GeoRL...5..411S,2005GeoRL..32.1104S,2005GeoRL..3221106S,2007MNRAS.381.1527J,2007PhRvL..98m1103C}.  No long-term modulation beyond 1 cycle is included in the model, and whether the solar cycles would increase or not is a result of the nonlinearity and stochasticity within the ongoing cycle \citep{Jiang2020}.  Such kind of model can produce the variation properties of solar cycles, such as the probability distribution of cycle amplitudes and even-odd rule \citep{2025ApJ...984..183W,2025RAA....25l5013W}.  Meanwhile, further prediction into the future is also not valid within the model, as every cycle is only related to its adjacent ones in either direction.

%In some occasions, a few ARs deviating from the Joy's law, coined as ``rogue active regions'', can prominently reduce or increase the polar field at cycle minimum and affect the strength of the next cycle \citep{2015ApJ...808L..28J,2017SoPh..292..167N}.

Some ARs, known as ``rogue active regions'', having low emerging latitude and tilt angle largely deviating from the Joy's law can have large impact on the polar field at cycle minimum \citep{2015ApJ...808L..28J,2017SoPh..292..167N}.  These ARs, if being anti-Joy, are also possible to significantly reduce the polar field.  However this is not our case, as the poleward surge we evaluate originates from a group of ARs that contribute a net increase to the polar field at cycle minimum, instead of some specific ARs with ``rogue'' properties.  This suggests that while being important part of the B-L mechanism, poleward surges, short-term polar field evolution, and the polar field at cycle minimum are not simply related by first inspirations.  A poleward surge with opposite polarity does not guarantee that the originating ARs contribute negatively to the final polar field, so the analysis of ARs and SFT processes is required to reveal their relationship.

%The probably weak cycle 26, under our understanding, is primarily of stochastic origin, instead of deterministic.  Form the current understanding of normal cycles (without different mechanisms for grand minima), it is not possible to answer the strength of further cycles -- we do not know if more upcoming cycles should be weak as well.  It is also possible for cycle 26 to fall into the range weaker than the lower limit considered in Jiang2020, which also means that the prediction could not extend to longer ranges. 

Our prediction model is based on the statistical properties of normal cycles, with the weakest amplitude being 107, without grand minima \citep{2018ApJ...863..159J,Jiang2020}.  The ongoing cycle is well within the normal cycle range, thus the prediction is well based on the model.  The next cycle 26 is also within the normal range considering the $1\sigma$ range, but may still be possible to be weaker than 107 if the $2\sigma$ range is considered.  From this perspective,  their is still a considerable probability for the model to be renewed in the future cycle.  Especially, the problem of the transition between normal cycles and grand minima is to be answered.

%If the polar field at cycle 25 and the amplitude of cycle 26 is indeed weaker than the previous range of normal cycles, grand minima may be possible if they have different mechanisms from normal cycles.
\begin{acknowledgements}

The international sunspot number version 2.0 is from the World Data Center SILSO, Royal Observatory of Belgium, Brussels, which is available at https://www.sidc.be/SILSO/datafiles.  The SDO/HMI data are courtesy of NASA/SDO and the HMI science teams. This research was supported by the National Natural Science Foundation of China through grant Nos. 12425305, 12350004, 12173005, 12503062, 12403067, 12373111, \& 12273061, the National Key R\&D Program of China through grant No. 2022YFF0503800, and supported by Specialized Research Fund for State Key Laboratory of Solar Activity and Space Weather.
\end{acknowledgements}

\bibliography{ref}{}
\bibliographystyle{aasjournal}
\end{document}